\documentclass[aps,pre,reprint,showpacs,notitlepage,onecolumn]{revtex4-1}
\usepackage{epsfig,graphics,amssymb,amsmath,subeqnarray,color,bm,mathtools,xcolor,float,graphicx}
\usepackage[toc,page]{appendix}
\usepackage[colorlinks=true,linkcolor=blue]{hyperref}

\numberwithin{equation}{section}
\definecolor{navyblue}{RGB}{0,0,128}
\definecolor{dodgerblue}{RGB}{30,144,255}
\definecolor{darkgrey}{RGB}{169,169,169}
\definecolor{deepskyblue}{RGB}{0, 191, 255}

\def\boldsymbol{\bm}

\def \veps{\varepsilon}

\def \bgamma{\boldsymbol{\gamma}}

\def \bgammad{\dot{\boldsymbol{\gamma}}}

\def \gammad{{\dot{\gamma}}}

\def \bnabla{\boldsymbol{\nabla}}

\def \bsigma{\boldsymbol{\sigma}}

\def \bzero{\mathbf{0}}

\begin{document}
\title{An active particle in a complex fluid}
\author{Charu Datt}
\affiliation{Department of Mechanical Engineering, 
University of British Columbia,
Vancouver, BC, V6T 1Z4, Canada
}
\author{Giovanniantonio Natale}
\affiliation{Department of Chemical and Biological Engineering, 
University of British Columbia,
Vancouver, BC, V6T 1Z3, Canada}

\author{Savvas G. Hatzikiriakos}
\affiliation{Department of Chemical and Biological Engineering, 
University of British Columbia,
Vancouver, BC, V6T 1Z3, Canada}

\author{Gwynn J. Elfring}\email{Electronic mail: gelfring@mech.ubc.ca}
\affiliation{Department of Mechanical Engineering, 
University of British Columbia,
Vancouver, BC, V6T 1Z4, Canada
}
\date{\today}

\begin{abstract}
In this work, we study active particles with prescribed surface velocities in non-Newtonian fluids. We employ the reciprocal theorem to obtain the velocity of an active spherical particle with an arbitrary axisymmetric slip-velocity in an otherwise quiescent second-order fluid. We then determine how the motion of a diffusiophoretic Janus particle is affected by complex fluid rheology, namely viscoelasticity and shear-thinning viscosity, compared to a Newtonian fluid, assuming a fixed slip-velocity. We find that a Janus particle may go faster or slower in a viscoelastic fluid, but is always slower in a shear-thinning fluid as compared to a Newtonian fluid.
\end{abstract}

\maketitle

\section{Introduction}
 Active particles are self-driven units which are capable of converting stored or ambient free-energy into systematic motion \citep{schweitzer2007brownian}.  These particles are found on length scales from subcellular to oceanic, and range from aquatic, terrestrial and aerial flocks to colloidal particles propelled through fluid by catalytic activity at their surface. The interactions of active particles with the medium they are found in, and amongst themselves, give rise to fascinating collective behaviour and beautiful pattern formation \citep{sriramSoft}. Active particles in fluid media can be either living, like swimming microorganisms \citep{eric_review}, or synthetic, like crystals of light-activated colloidal surfers \citep{livingcrystal},  swimming droplets \citep{shashi} and chemically self-propelled nano-motors \citep{kapral_nano}. For sufficiently small sizes of active particles, inertial forces are negligible compared to viscous forces, and one may assume the fluid to be under an instantaneous equilibrium of forces \citep{purcell}.
 
Several microorganisms propel themselves using small surface distortions like in the coordinated beating of cilia on \textit{Opalina} and \textit{Paramecium}  \citep{sleigh2016biology}. As such, these swimmers are often modelled as spheres with a prescribed surface slip-velocity \citep{Pedley_review}; the slip-velocity serves as a coarse-grained description of any deformation or dynamics on the particle body that leads to its motion \citep{Lighthill1951, BlakeJFM}. Likewise, a chemically active colloidal particle with asymmetric catalytic properties generates a non-uniform distribution of reaction products and hence, also a flow within a thin `inner' region near the particle's surface \citep{anderson1989colloid}. The surface flow  and the resultant diffusiophoretic motion may also be modelled by prescribing an apparent slip-velocity on the particle surface \citep{prost}. The motion of these particles, arising due to a surface slip-velocity is, by now, well-understood for particles  that move in Newtonian fluids at low Reynolds numbers \citep{brennen, elgeti_review}. In general, the propulsive force generated by the surface slip-velocity balances the hydrodynamic drag force due to the rigid body motion of the particle. For simple bodies, the swimming velocity is given directly by the surface average of the prescribed slip-velocity \citep{GwynnReci} and because of this simplification, detailed models of the surface slip-velocity for living and synthetic active particles are often unnecessary.

In contrast, an understanding of dynamics of active particles in non-Newtonian fluids is still developing \citep{arratia_review}. Unlike in Newtonian fluids, the constitutive equation for stress is nonlinear in non-Newtonian fluids and as a result a straightforward linear decomposition of the flow field into drag and thrust components fails \citep{jfmCharu}. Consequently, a surface average of the slip-velocity does not yield the velocity of the particle, and so a detailed description of the surface slip velocity may be significant in complex fluids. Despite this, many recent studies consider, as a point of comparison with Newtonian fluids, the `two-mode' swimmer model \citep{ Maffettone,Li2014, montenegro, Lailai}, although recently it was shown that neglected details of the surface slip-velocity can have a qualitative effect on the motion of the particle in a shear-thinning fluid \citep{jfmCharu}.

In this work, we analyse the motion of an active particle in a weakly nonlinear complex fluid with a \textit{general} axisymmetric slip velocity by means of the reciprocal theorem \citep{stone_samuel,Lauga_theorem}. This allows us to consider a complete range of prescribed motions on the particle surface and to determine what details matter and why. We note that the swimming gait (apparent surface slip-velocity) of the swimmer may itself be affected in complex fluids as compared to Newtonian fluids, due to, for example, constraints on power for biological swimmers, or changes in solute diffusivity for diffusiophoretic particles. Here, however, we consider swimmers with the same swimming gait as in Newtonian fluids. As an example, we consider the slip velocity of self-diffusiophoretic `Janus' particles and discuss the effects of viscoelasticity and shear-thinning rheology on the particles' propulsion velocity. These active colloidal particles, at times, may swim through polymer suspensions \citep{New_Bechinger}, and an understanding of their dynamics in complex fluids may lead to interesting applications in biological and chemical engineering \citep{Popescu2016}. Recent studies on the effects of rheology on the motion of Janus particles \citep{GomezPrl, PRFStone}  have shown that the particle translational and rotational dynamics are coupled in media with viscoelasticity or local viscosity-variations. Further, motivated by recent works on the dynamics of active particles in background flow of non-Newtonian fluids \citep{ardekani_background, yeoman_background, decorato}, we generalize the reciprocal theorem formulation \citep{Eric2016, elfring2015theory, Lauga_theorem} to include a background flow in the spirit of previous classical work on passive particles in weakly nonlinear flows \citep{Leal_review}.

\section{Modelling active particles}
\label{active_particle}
Biological microswimmers possess variety of different geometries and swimming modes; many, like ciliates (\textit{Opalina}) and multicellular colonies of flagellates (\textit{Volvox}), are approximately spherical in shape and propel due to the beating of closely packed cilia on their surface \citep{sleigh2016biology}. These swimmers, in an idealised model, are mathematically represented as spheres with small amplitude radial and tangential motions of elements of the surface. The original model (now known as the squirmer model), by  \citet{Lighthill1951} and \citet{BlakeJFM}, considered only axisymmetric surface distortions so the swimmers could swim only along their axis of symmetry. Recently, \citet{Pak2014} extended the model to arbitrary surface deformations allowing three-dimensional translational and rotational swimming kinematics of the swimmer.

Synthetic active particles, too, can be conceived in many shapes with a variety of propulsion mechanisms \citep{Janus_review}. Self-phoretic particles, in particular, are colloids which are able to generate local gradients through the catalytic physiochemical properties on their surface \citep{Golestanian2, Ajdari-Gole, EricMichelin}. The short-range interaction between the surface of the swimmer and the self-generated outer field-gradient (solute concentration, temperature or electric field) locally creates fluid motion in the vicinity of particle boundary   that leads to particle propulsion due to phoresis \citep{anderson1989colloid}. When the interaction layer is thin compared to the particle size, phoretic effects can be represented by the generation of slip-velocities on the particle surface \citep{prost, EricMichelin}.

In this work, we focus on spherical phoretic particles \citep{Ajdari-Gole, EricMichelin}, with an axisymmetric slip velocity expressed here as a series of Legendre polynomials
\begin{align}
\bm{u}^S\left(\theta,t\right) = \sum_{p=1}^{\infty} \alpha_p \left( t\right)K_p \left( \cos{\theta} \right)\ \bm{\hat{e}}_{\theta} 
\label{slip}
\end{align}
with
\begin{align}
 K_p \left( \cos{\theta} \right) = \frac{(2p+1)}{p\left(p+1\right)}P_p^{'} \left(\cos{\theta}\right)  \sin{\theta},
 \end{align}
where $\theta$ is the polar angle and $P_p$ is the $p^{\text{th}}$ Legendre polynomial \citep{sebastien_nutrient}. The flow field due to the swimmer in Newtonian fluids is completely characterised and determined by the intensities of the `squirming' modes, $\alpha_p$ \citep{BlakeJFM}.  Of particular significance are the first two modes: $\alpha_1$, which fixes the swimming velocity \citep{Lighthill1951}, and $\alpha_2$,  which defines the strength of the force dipole generated by the swimmer $\Sigma = 10\pi \alpha_2$ \citep{EricMichelin}. Consequently, for analyses of collective behaviour \citep{Stark, delfau_collective}, or transport of nutrients \citep{sebastien_nutrient, ishikawa_nutrient}, in Newtonian fluids, active particles are very often modelled with a truncated slip-velocity expansion which retains only the first two terms. We consider here only steady slip-velocities on the particle surface, which is often appropriate for phoretic particles; however, in general, especially for models of biological organisms where the surface motion arises from a cyclical deformation, the slip velocities may depend on time \citep{pedley_with_swirl}. This time dependence of the surface actuation is then particularly important for fluids which possess history dependence, like polymer solutions, especially when the time scale of surface actuation is of the same order as the fluid relaxation time \citep{elfring16}. 

Self-diffusiophoretic particles propel due to asymmetric surface chemical reactions \citep{anderson1989colloid, Golestanian2, brady2011} which cause an induced imbalance of osmotic effects in a thin interaction layer on the particle surface. The resulting flow in this thin layer, the apparent slip-velocity, is proportional to the local solute concentration gradient and the specifics of solute-surface interactions (phoretic mobility). Under the assumption that diffusion is fast enough so that the chemical reaction at the surface is controlled by the far-field solute concentration (fixed-flux formulation, D\"{a}mkohler number = 0) and on neglecting the distortion of solute distribution due to flow resulting from phoretic effects (P\'{e}clet number = 0), one obtains the squirming modes in \eqref{slip}
\begin{align}
\alpha_p = \frac{pA_p}{2p+1} \frac{M}{D}, 
\label{squirm_mode}
\end{align}
where the surface activity $A\left(\theta\right) = \sum A_p P_p \left( \cos{\theta}\right)$ (and positive values denote absorption of solute), the phoretic mobility $M$ is assumed to be constant over the surface and $D$ is the solute diffusivity (see \citet{EricMichelin} for details).

\begin{figure}
\center
\includegraphics[width = 0.25\textwidth]{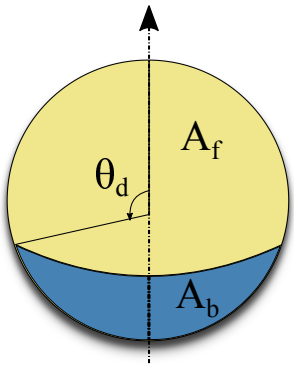}
\caption{Self-phoretic particle with two compartments of different activity, $A_f$ and $A_b$. We consider particles with a constant uniform mobility over the surface. When $\theta_d = \pi/2$, the particle has compartments of equal cover, which we call a symmetric Janus particle.   }
\label{janus-fig}
\end{figure}

We consider Janus type particles with a discontinuous change in activity between two distinct compartments of the surface activity, $A(\theta) = A_f$ for $\theta < \theta_d$ while $A(\theta) = A_b$ for $\theta > \theta_d$ as illustrated in figure \ref{janus-fig}. Here, we take the rear compartment to be inert, $A_b = 0$, in which case the coefficients are given by \citep{EricMichelin}
\begin{align}
A_0 =\frac{ A_f}{2} \left(1-\cos{\theta_d}\right),  \quad A_n =  \frac{A_f}{2} \left[ P_{n-1} (\cos{\theta_d}) - P_{n+1} (\cos{\theta_d})\right] \quad (n\geq 1), 
\label{michelin_eq}
\end{align}
which then set the squirming modes and the entire flow field for Janus particles  in Newtonian fluids.

\section{Swimming in background flow of weakly non-Newtonian fluid }
Consider a general active particle (or swimmer) $\mathcal{B}$ with surface $\partial{\mathcal{B}}$ immersed in a background flow $\bm{u}^{\infty}$ of an incompressible and weakly nonlinear complex fluid. The velocity on the swimmer surface $\partial{\mathcal{B}}$ is
\begin{align}
 \bm{u} \left( \bm{x} \in \partial \mathcal{B} \right) = \bm{U} + \bm{\Omega} \times \bm{x} + \bm{u}^S,
\end{align}
where $\bm{U}$ is the translational velocity of the particle, $\bm{\Omega}$ is the rotational velocity and $\bm{u}^S$ is the prescribed deformation velocity on its surface (the swimming gait).

The rheology of the non-Newtonian fluid is assumed to be only weakly nonlinear \citep{ Lauga_theorem, elfring2015theory}, and thus, a constitutive equation of the form
\begin{align}
\bm{\tau} = \eta \dot{\bm{\gamma}} + \varepsilon \bm{\mathsf{A}}[\bm{u}], \label{generic_eq}
\end{align} 
where $\bm{\tau}$ is the deviatoric stress, $\eta$ is the viscosity and $\dot{\bm{\gamma}}$ the strain-rate tensor such that $\eta \dot{\bm{\gamma}}$ gives the Newtonian contribution. $\bm{\mathsf{A}}[\bm{u}]$ is a symmetric tensor and a nonlinear functional of $\bm{u}$ and $\veps$ is a small dimensionless parameter characterising the deviation from Newtonian behaviour, for example, small Deborah number in case of viscoelastic fluids, or small Carreau number for shear-thinning fluids.

We consider the flow field to be inertialess and in mechanical equilibrium with $\bnabla\cdot\bsigma=\bzero$, where $\bm{\sigma}$ is the stress tensor corresponding to the velocity field $\bm{u}$. We define disturbance fields $\bm{u}' = \bm{u} - \bm{u}^{\infty}$ and $\bm{\sigma}' = \bm{\sigma} - \bm{\sigma}^{\infty}$ where $\bm{u}^{\infty}$ and $\bm{\sigma}^{\infty} $ correspond to the velocity and stress fields of the background flow in the absence of the particle. Due to the nonlinearity of constitutive equation (\ref{generic_eq}),  $\bm{u}'$ and $\bm{\sigma}'$, in general, do not represent velocity and stress fields of the same problem (except when $\veps = 0$).

\citet{stone_samuel} demonstrated a shortcut to obtain the swimming velocity of an arbitrary swimmer in a Newtonian fluid with a given prescribed surface actuation $\mathbf{u}^S$ without calculation of its unknown flow field using the Lorenz reciprocal theorem in low-Reynolds-number hydrodynamics \citep{happel2012low}, provided one can solve the rigid body resistance/mobility problem for a body of the same shape.  Using this approach \citet{deborah_number, Lauga_theorem} then developed integral theorems to determine the swimming velocity in complex fluids. We use these methods below, following the formulation in \citep{elfring2015theory, elfring16}, to obtain the swimming velocity of a swimmer in a weakly non-Newtonian fluid but include the possiblity of a non-zero background flow for generality. 

For the resistance problem (denoted with a hat), we consider rigid body motion with translational velocity $\hat{\bm{U}}$ and rotational velocity $\hat{\bm{\Omega}}$, through a Newtonian fluid with corresponding velocity field $\bm{\hat{u}}$ and associated stress tensor $\bm{\hat{\sigma}}=\hat{\eta}\hat{\dot{\bm{\gamma}}}$.  As both flows (due to the swimmer and due to rigid-body motion) are in mechanical equilibrium, we have
\begin{align}
\begin{aligned}
&\hat{\bm{u}} \cdot \left(\nabla \cdot \bm{\sigma}' \right)= \bm{u}' \cdot \left( \nabla \cdot \hat{\bm{\sigma}}\right) = 0.
\end{aligned}
\end{align}
Integrating over the volume of fluid, $\mathcal{V}$, exterior to $\mathcal{B}$ and applying the divergence theorem while enforcing the incompressiblity of the flows, we get 
\begin{align}
\int_{\partial \mathcal{V}} \bm{n} \cdot \bm{\sigma}' \cdot \hat{\bm{u}} \ dS + \int_{\mathcal{V}} \bm{\tau}' : \nabla \hat{\bm{u}} \ dV = \int_{\partial \mathcal{V}} \bm{n} \cdot \hat{\bm{\sigma}} \cdot \bm{u}' \ dS + \int_{\mathcal{V}} \hat{\bm{\tau}} : \nabla \bm{u}' \ dV = 0,
\label{reci_eq}
\end{align}
where we have defined $\bm{\tau}' = \eta \dot{\bm{\gamma}}' + \varepsilon \bm{\mathsf{A}}'$ and $\bm{\mathsf{A}}' = \bm{\mathsf{A}}\left[ \bm{u}\right] - \bm{\mathsf{A}}\left[ \bm{u}^{\infty}\right]$. 
The surface $\partial \mathcal{V}$ that bounds the fluid volume $\mathcal{V}$ is composed of the body surface, $\partial \mathcal{B}$, and an outer surface (fluid or solid, possibly at infinity). Here, $\bm{n}$ is the normal to the surface, $\partial \mathcal{V}$, pointing into $\mathcal{V}$.

Provided the fields, $\bm{u}'$ and $\bm{\sigma}'$, decay appropriately in the far-field, we may neglect the outer surface of $\partial\mathcal{V}$ (we shall show this is the case for weakly viscoelastic linear background flows in a subsequent work). For flows bounded by no-slip walls these terms will be identically zero. Upon substitution of the boundary conditions on $\partial\mathcal{B}$ for each field and enforcing that the net hydrodynamic force, $\bm{F}=\int_{\partial\mathcal{B}}\bm{n}\cdot\bm{\sigma}\ dS$ , and torque, $\bm{L}=\int_{\partial\mathcal{B}}\bm{x}\times(\bm{n}\cdot\bm{\sigma})\ dS$, are both zero on a free swimmer in the absence of inertia, the left-hand side of \eqref{reci_eq} simplifies to 
\begin{align}
\eta \int_{\mathcal{V}} \dot{\bm{\gamma}}' : \nabla \hat{\bm{u}} \ dV + \varepsilon \int_{\mathcal{V}} \bm{\mathsf{A}}' : \nabla \hat{\bm{u}} \ dV = 0.
\label{lhs}
\end{align}
while the right-hand side of \eqref{reci_eq} simplifies to 
\begin{align}
\hat{\bm{F}} \cdot \bm{U} + \hat{\bm{L}} \cdot \bm{\Omega} + \int_{\partial \mathcal{B}} \bm{n} \cdot \hat{\bm{\sigma}}\cdot \left( \bm{u}^S - \bm{u}^{\infty}\right) \ dS + \hat{\eta}\int_{\mathcal{V}}\dot{\bm{\gamma}}' : \nabla \hat{\bm{u}} \ dV = 0,
\label{rhs}
\end{align}
where we have utilized the fact that $\hat{\dot{\bm{\gamma}}} : \nabla \bm{u}' = \dot{\bm{\gamma}}' : \nabla \hat{\bm{u}}$. We will here use 6-dimensional vectors for compactness, $\bm{\mathsf{U}} = [\bm{U} \quad \bm{\Omega} ]^T$ and $\hat{\bm{\mathsf{F}}} = [\hat{\bm{F}} \quad \hat{\bm{L}}]^T$, and from the linearity of the Stokes equation,  write  $\hat{\bm{u}} = \hat{\bm{\mathsf{L}}} \cdot \hat{\bm{\mathsf{U}}}$, $\bm{\hat{\sigma}} = \hat{\bm{\mathsf{T}}} \cdot \hat{\bm{\mathsf{U}}}$ and $\hat{\bm{\mathsf{F}}} = -\hat{\bm{\mathsf{R}}} \cdot \hat{\bm{\mathsf{U}}}$, where $\hat{\bm{\mathsf{R}}}$ is symmetric.  Finally, upon combining \eqref{lhs} with \eqref{rhs} we obtain
\begin{align}
\bm{\mathsf{U}} = \hat{\bm{\mathsf{R}}}^{-1} \cdot \left[ \int_{\partial \mathcal{B}} \left( \bm{u}^S - \bm{u}^{\infty}\right) \cdot \left( \bm{n} \cdot \hat{\bm{\mathsf{T}}}\right) \ dS - \varepsilon \frac{\hat{\eta}}{\eta} \int_{\mathcal{V}} \bm{\mathsf{A}}' : \nabla \hat{\bm{\mathsf{L}}} \ dV     \right], 
\label{propulsion_vel}
\end{align}
which gives us a relation for the propulsion velocity of a swimmer in the background flow of a weakly non-Newtonian fluid. The correction to the Newtonian swimming speed, due to the tensor $\bm{\mathsf{A}}'$, typically depends upon the unknown field $\bm{u}$ but, upon expanding perturbatively in $\varepsilon$, the correction depends only on the Newtonian solution to leading order.

For a spherical particle of radius $a$, the translational velocity is given simply by
\begin{align}
\bm{U} = - \frac{1}{4\pi a^2} \int_S  \left( \bm{u}^S - \bm{u}^{\infty} \right)dS - \varepsilon \frac{1}{8\pi \eta} \int_{\mathcal{V}} \bm{\mathsf{A}}' : \left( 1+ \frac{a^2}{6} \nabla^2 \right) \nabla \bm{\mathsf{G}}\  dV
\label{for_sphere}
\end{align}
where $\bm{\mathsf{G}} = \left(\bm{\mathsf{I}} + \bm{r}\bm{r}/r^2 \right)/r$ is the Oseen tensor (or Stokeslet).  As expected, when $\varepsilon = 0$, one obtains the result for a swimmer in a background flow of Newtonian fluid \citep{GwynnReci}.

\section{Janus particle in non-Newtonian fluids}

As examples of an active particle in a complex fluid, we study a Janus particle in a weakly viscoelastic fluid and in a weakly shear-thinning fluid but assume the same surface slip-velocity as in the Newtonian fluid (given by \eqref{squirm_mode}). We note that we expect the non-Newtonian rheology will also affect the slip velocity for phoretic particles but focus here only on kinematic differences for a fixed swimming gait. Viscoelasticity and shear-thinning rheology are two important non-Newtonian properties \citep{bird1987dynamics} and also the characteristics of many biological fluids \citep{merrill_blood, Lai_mucus} wherein these artificial swimmers have potential applications \citep{Nanodreams}.  As discussed in \S \ref{active_particle}, we assume the diffusion of the solute to be fast enough so that the effects of P\'{e}clet and Damk\"{o}hler number can be neglected and we shall consider the particle in an unbounded and otherwise quiescent background ($\bm{u}^\infty=\bzero$). We first analyse the Janus particle in a weakly viscoelastic fluid.

\subsection{Viscoelasticity: second-order fluid}
Viscoelastic fluids exhibit both viscous and elastic responses to forces. Such fluids possess a memory, and stresses in them depend on the flow history. For flows which are both slow and slowly varying, viscoelasticity may be modelled without any memory of the past stresses as a second-order fluid \citep{morozov2015introduction},
 \begin{align}
 \bm{\tau} = \eta \bgammad - \frac{\Psi_1}{2} \buildrel \nabla \over {\bgammad} + \Psi_2 \bgammad \cdot \bgammad.
 \end{align}
 Here, $\eta$ is the total viscosity of the solution, and $\Psi_1$ and $\Psi_2$ are the first and second normal stress-difference coefficients, respectively. The first normal stress difference is generally positive in viscoelastic flows i.e  $\Psi_1 > 0$. The triangle denotes the upper-convected derivative
 \begin{align}
 \buildrel \nabla \over \bgammad = \frac{\partial \bgammad}{\partial t} + \bm{u} \cdot \nabla \bgammad - \left( \nabla \bm{u}\right)^T \cdot \bgammad - \bgammad \cdot \nabla \bm{u}.
 \end{align}
 
In order to study the effect of fluid rheology on the particle, we first non-dimensionalise the equations by scaling lengths with the particle radius, $a$; velocities with the first swimming mode $ \alpha_1$, which without any loss of generality is assumed to be positive,  and  stresses with $\eta \omega$, where $\omega =  \alpha_1 / a$ is the scale of strain-rate. The resulting dimensionless constitutive equation is
  \begin{align}
 \bm{\tau}^* = \bgammad^* - De \left( {\buildrel  \nabla \over {\bgammad^*}}  + b \bgammad^* \cdot \bgammad^* \right),
 \end{align}
 with $De = \omega \Psi_1/2 \eta$, the Deborah number, which is the ratio of the relaxation time scale of the fluid to the characteristic timescale of the flow and $b = -2\Psi_2/ \Psi_1 \ge 0$. Henceforth, we work in dimensionless quantities and drop the stars (*) for the sake of convenience. For small $De$ (weakly viscoelastic limit), we expand the flow quantities in a regular perturbation expansion in $De$ \citep{propulsion_eric, elfring2015theory, Maffettone} to get, at the leading order,
  \begin{align}
 \bm{\tau}_0 = \bgammad_0,
 \end{align}
 and at $O\left( De\right)$
 \begin{align}
 \bm{\tau}_1 = \bgammad_1 +\bm{\mathsf{A}},
 \label{expansion_de}
 \end{align}
 with $\bm{\mathsf{A}} = - \left(\buildrel \nabla \over {\bgammad_0} + b \bgammad_0 \cdot \bgammad_0  \right)$.
 The angular velocity of  a spherical swimmer is zero due to axisymmetry while its translational velocity, correct to $O\left( De\right)$, is given by \eqref{for_sphere} where now $\epsilon = De$.

The flow field for a swimmer with prescribed surface velocity \eqref{slip} in a quiescent Newtonian fluid is given by \citep[see][]{ishikawa}
\begin{align}
\begin{aligned}
\bm{u}_0 = &-\frac{1}{2r^3} \bm{e} + \frac{3}{2r^3} \frac{\bm{e}\cdot\bm{r}}{r}\frac{\bm{r}}{r} + \sum_{p=2}^{\infty}\left( \frac{1}{r^{p+2}} - \frac{1}{r^p}\right) \left( p + \frac{1}{2}\right)\Theta_p P_p \left( \frac{\bm{e}\cdot \bm{r}}{r}\right) \frac{\bm{r}}{r} \\
& + \sum_{p= 2}^{\infty} \left( \frac{p}{2r^{p+2}} - \left(\frac{p}{2}-1\right) \frac{1}{r^p}\right) \left(p + \frac{1}{2} \right)\Theta_p W_p \left(\frac{\bm{e}\cdot\bm{r}}{r}\right) \left(\frac{\bm{e}\cdot\bm{r}}{r} \frac{\bm{r}}{r} - \bm{e} \right),
\end{aligned}
\end{align} 
where $\bm{e}$ is the swimming direction and $\bm{r}$ is the position vector with $r = \vert \bm{r}\vert$ from the centre of the sphere. $\Theta_p = \alpha_p/ \alpha_1$ and $W_p\left(x\right) = 2/\left(n\left(n+1\right) \right)P'_p \left(x\right)$. Using the Newtonian velocity field, one can calculate the strain-rate field around the swimmer, $\dot{\bgamma_0}$, and thus obtain the expression for $\bm{\mathsf{A}}$. Substituting the expression for $\bm{\mathsf{A}}$ in \eqref{for_sphere} and using the orthogonal properties of Legendre polynomials, one obtains, after some lengthy but straightforward calculations,
 \begin{align}
U / U_N= 1 +  De\left(b-1\right)\sum_{p=1}^{\infty} C_p\Theta_p\Theta_{p+1},
\label{equation_vel}
\end{align}
where 
\begin{align}
C_p = \dfrac{6p }{(p+1)^2(p+2)}.
\end{align}
Recall that $U_N = \alpha_1$ is the (dimensional) swimming speed in Newtonian fluids. Frequently, the slip-velocity description is truncated at two modes i.e. \mbox{$\Theta_p = 0 \: \forall \: p> 2$}, and depending on whether $\Theta_2 < 0$, $\Theta_2=0$ or $\Theta_2>0$ the swimmer is identified as a pusher, neutral or puller swimmer, respectively, in Newtonian fluids \citep{elgeti_review}. However, swimmers like starfish larvae \citep{Manu_new}, and Janus particles possess significant values of higher modes. When considering such swimmers in non-Newtonian fluids, one should be careful while truncating the series because unlike in Newtonian fluids, swimming speeds may be qualitatively affected by higher modes \citep{jfmCharu}. Indeed, as can be noted from \eqref{equation_vel}, setting the modes $\alpha_1 = 1$, $\alpha_2 = 1$ and $\alpha_3 = 2$  (with appropriate units) produces qualitatively different swimming behaviour than $\alpha_1 = 1$, $\alpha_2 = 1$ and $\alpha_3 = -2$ when just the first three modes are considered. Therefore, the expression \eqref{equation_vel}, while giving the contribution of all spectral modes in the slip-velocity expansion to the swimming velocity, helps to predict when it may be reasonable to neglect higher modes and use a simple `two-mode' description to obtain the swimming speed.

We consider the case of symmetric Janus particle, where precisely one half is chemically active and the other inert, $\theta_d = \pi/2$. The spectral coefficients for activity in this case are zero for even modes (from \eqref{michelin_eq}), and consequently $\Theta_{2p} = 0$. Hence, from \eqref{equation_vel}, one finds that a symmetric Janus particle  (with a constant uniform surface mobility) swims only at its Newtonian speed -- a result also true for a two-mode neutral swimmer  \citep{Maffettone}  but here obtained without any restriction on the number of modes being considered. Interestingly, one could obtain this result by observing that the non-Newtonian contribution in \eqref{for_sphere} is a volume integral of the contraction of an even tensor $\bm{\mathsf{A}}$ (under $\bm{x}\rightarrow-\bm{x}$) and an odd kernel and therefore vanishes. Similarly, looking at the power consumption of a squirmer, $P$, correct to the first order \citep{Maffettone} 
\begin{align}
2P =   \int_{\mathcal{V}} \dot{\bm{\gamma}}_0 : \dot{\bm{\gamma}}_0 \ dV + De \int_{\mathcal{V}} \bm{\mathsf{A}}: \dot{\bm{\gamma}}_0 \ dV,
\end{align}
one finds once again that for a symmetric Janus particle the non-Newtonian contribution gives a null result. Thus, a symmetric Janus particle in a second-order fluid swims and expends power as if in an equivalent Newtonian fluid ($De=0$), correct to the first order in $De$, for the same surface slip-velocity as in the Newtonian fluid. We note that the non-Newtonian rheology will affect the solution of the `inner' region for phoretic particles \citep{EricMichelin}. Additional non-Newtonian stresses arise on the particle surface, and even the solute diffusivity may change due to viscosity variations. For a thin interaction layer, neglecting effects of P\'{e}clet and Damk\"{o}hler number, the slip velocity will change at $O\left( De\right)$ similarly to the case of electrophoresis considered by \citet{adityakhair_PRE}. Here, however, our emphasis is on studying the changes in the propulsion velocity from its Newtonian value for a given (but arbitrary) slip velocity on the particle surface. 

A similar result was obtained by \citet{Leal-fore-aft} for axisymmetric passive particles with fore-aft symmetry in a second-order fluid, where such particles translate, to the first approximation, at the same rate as in an equivalent Newtonian fluid. On comparison with present results, one may expect even non-spherical active particles with fore-aft symmetry in second-order fluids to behave as if in equivalent Newtonian fluids.

\begin{figure}
\center
\includegraphics[width = 0.7\textwidth]{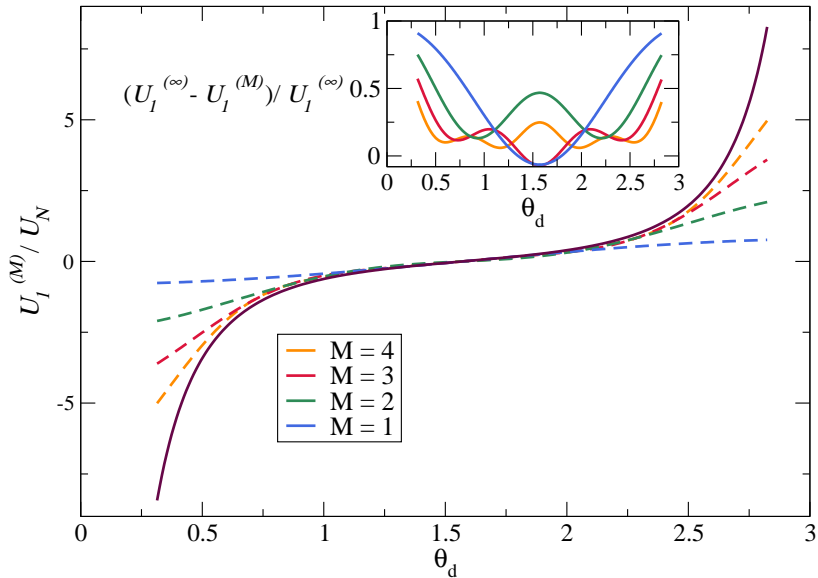}
\caption{Variation of the scaled first-order swimming velocity $U_1^{\left( M\right)}/ U_N$ with $\theta_d$ obtained for the first $M+1$ modes (dashed lines), and $b = 0.2$. $U_1^{\left({\infty}\right)}$ corresponds to the convergence value ($M = 99$) and is depicted by the solid line. Inset plot shows the relative error.}
\label{ve_contri}
\end{figure}

When the two halves of the Janus particle are not exactly equal, i.e. $\theta_d \neq \pi/2$, then the even spectral modes of the activity, $A_{2p}$, are no longer equal to zero, and hence $\Theta_{2p} \neq 0$. Consequently, the non-Newtonian contribution to the swimming velocity may now be non-zero, and can be easily calculated for any level of active surface coverage, $\theta_d$. We find that when $\theta_d > \pi/2$, the particle swims faster than in a Newtonian fluid and while for $\theta_d < \pi/2$ it swims slower, provided $b< 1$ (see \cite{christov16} and \cite{decorato16b} for a recent discussion on permissible values of $b$). Interestingly, one can qualitatively predict this result by considering the two-mode description, by observing that $\Theta_2 = 2 \cos{\theta_d}$. The former particle behaves as a pusher, $\Theta_2 < 0$, and thus swims faster, where as the latter is a puller, $\Theta_2 > 0$, and therefore swims slower than in a Newtonian fluid (from \eqref{equation_vel}), as also reported for two-mode swimmers by \citet{Maffettone}. Quantitatively, the viscoelastic contribution decays for higher modes as $C_p \sim 1/p^2$ and a two-mode description gives the viscoelastic contribution with a relative error of less than 0.1 for $\vert \cos{\theta_d} \vert \leq 0.35$; however, the approximation grows worse upon increasing the fore-aft asymmetry of the particle and a three-mode description is better for $\vert \cos{\theta_d} \vert \geq 1/\sqrt{5}$. This is shown in figure \ref{ve_contri}, where we plot the scaled first-order velocity, $U_1^{\left(M\right)}/U_N = \left( b-1\right)\sum_{p=1}^{M} C_p\Theta_p\Theta_{p+1}$ from \eqref{equation_vel} for different coverage areas of activity with varying number of modes. Note that as $\theta_d$ approaches $0$ or $\pi$, the Newtonian velocity $U_N \rightarrow 0$ and $U_1^{\left(M\right)}/U_N$ diverges.

The asymptotic results for a small $De$ expansion are seen to be valid for only very small values of $De$ ($\approx 0.02$ for two-mode swimmers with $O\left(1\right)$ modes \citep{Maffettone}). This may be understood by noting that squirming modes of magnitude $O\left( 1\right)$ result in strain-rates of magnitude  $O(10)$ on the surface of the particle in a Newtonian fluid and, therefore, $O(10^2)$ values of the non-Newtonian contribution $\bm{\mathsf{A}}$, which thereby renders the Deborah number expansion accurate for only very small values of $De$.  Numerical results using the Giesekus model, at higher values of $De$, find all swimmers -- pusher, puller and neutral -- swimming slower and expending less power than in an equivalent Newtonian fluid \citep{Lailai}; although, one might expect results obtained using the second-order fluid model to deviate from those obtained with the Giesekus model, at moderate Deborah numbers, due to the saturation of polymer elongation in the latter and the associated differences in extensional rheology. In the experimental study of Janus particles in viscoelastic fluids by \citet{GomezPrl}, the Deborah (Weissenberg) numbers were quite small, and hence in a regime where one may then expect the second-order model to, at least qualitatively, predict the viscoelastic fluid behaviour \citep{leal79}.

\subsection{Shear-thinning rheology: Carreau model}
\begin{figure}
\center
\includegraphics[width = 0.7\textwidth]{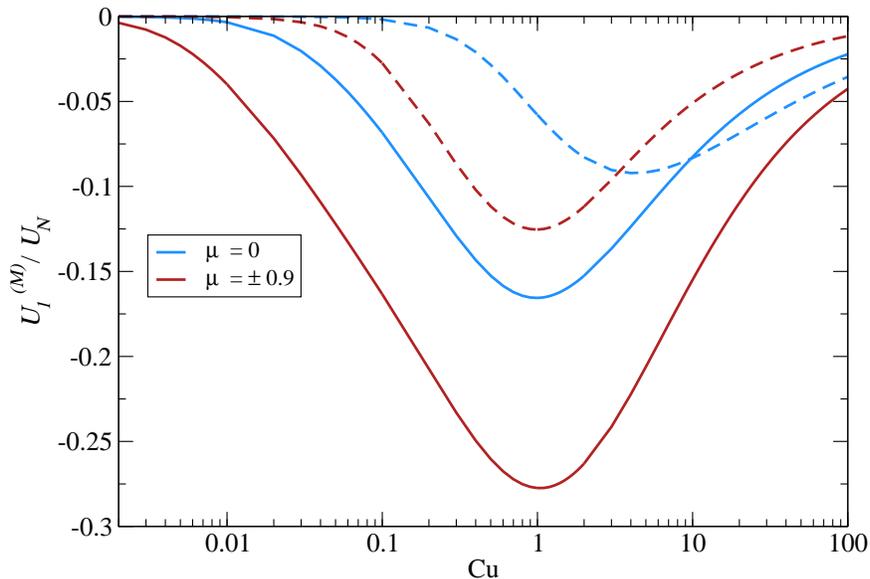}
\caption{Variation of the scaled first-order swimming velocity $U_1^{\left( M\right)}/U_N$ (obtained for $M+1$ modes) with $Cu$ for two values of $\mu \equiv \cos{\theta_d}$. Solid lines correspond to $M = 30$, and $M = 28$ for $\mu = 0$ (symmetric) and  $\mu = \pm 0.9$ respectively (additional modes lead to negligible differences). Dashed-lines correspond to the swimming velocity with just the first two modes.}
\label{Shearthinning_contri}
\end{figure}

Shear-thinning fluids experience a loss of apparent viscosity with applied strain-rate. The Carreau-model \citep{bird1987dynamics} and its perturbation to the form in (\ref{generic_eq}) has recently been covered by \citet{jfmCharu}. We consider the perturbation of  the flow quantities in the viscosity ratio, $\varepsilon = 1 -\beta$ where $\beta \in \left[ 0,1\right]$ is the ratio of infinite shear-rate viscosity to zero shear-rate viscosity, as this expansion is uniformly valid for all strain rates and obtain $\bm{\mathsf{A}} = \left\{ -1 + \left( 1+ Cu^2 \vert \gammad_0\vert ^2\right)^{\left(n-1\right)/2} \right\}\bgammad_0$.  Here, $Cu$, the Carreau number is the ratio of the characteristic strain-rate in the flow, to the cross-over strain-rate in the fluid and $n$ characterizes the degree of shear-thinning ($n<1$). With this form of $\bm{\mathsf{A}}$, it is difficult to obtain an analytical expression for the propulsion velocity similar to that obtained for the viscoelastic case \eqref{equation_vel}. However, one can numerically calculate the propulsion velocity with higher modes and then compare the results with just the first two modes. This is done in figure \ref{Shearthinning_contri} for $n = 0.25$, where we plot $U_1^{\left(M\right)}/U_N$ for two values of $\mu \equiv \cos{\theta_d}$.

We find that irrespective of the position of $\theta_d$, the Janus particle swims slower in a shear-thinning fluid than in a Newtonian fluid. The non-monotonic variation of the first-order swimming speed with $Cu$ in figure \ref{Shearthinning_contri} is similar to as found by \citet{jfmCharu} for any two-mode squirmer. Though the two-mode description qualitatively predicts the results: all -- neutral, pusher and puller -- swimmers swim slower, with pusher and pullers swimming at the same velocity \citep{jfmCharu}, it is apparent from figure \ref{Shearthinning_contri} that higher modes may significantly alter the results.  Additionally, we note that the values of $\Theta_2$ and $\Theta_3$ for any Janus particle lie in the range where \citet{jfmCharu} predict a smaller swimming velocity than in Newtonian fluids.

\section{Conclusion and future work}
In this work, we studied active particles with prescribed surface velocities in non-Newtonian fluids. Using the reciprocal theorem, we derived a general form of the propulsion velocity of an active particle in a weakly nonlinear background flow. Using this formalism, we calculated the swimming speed for an active particle with a general, axisymmetric slip velocity in an otherwise quiescent second-order fluid extending results previously obtained for a two-mode description. We then considered the motion of diffusiophoretic Janus particles in weakly viscoelastic and shear-thinning fluids. We showed that a Janus particle with two equal halves, in a weakly viscoelastic fluid, will swim at the same speed as in a Newtonian fluid due to its fore-aft symmetry (provided the surface slip-velocity remains unchanged). When this symmetry is broken the particle may swim faster or slower than in a Newtonian fluid and this may be predicted by considering the Janus particle as a pusher or puller based on the two-mode squirmer description. Conversely, in a weakly shear-thinning fluid, a Janus particle always swims slower than in a Newtonian fluid.

While analysing Janus particles, we neglected any changes to the slip-velocity due to fluid rheology as well any dynamics due to the distortion of the solute concentration field of phoretic particles because of the velocity field. The latter may not be true for large proteins or molecules, when the diffusion constant is small and the P\'{e}clet number becomes significant. This coupling of the velocity and concentration field leads to interesting dynamics in Newtonian fluids \citep{Michelinauto,EricMichelin} and is an avenue for further inquiry in non-Newtonian fluids. We also expect the fluid rheology to affect the slip-velocity of an active particle: the gait of a biological microswimmer may be modified by non-Newtonian stresses, likewise the slip-velocity of a diffusiophoretic Janus particle. For a complete understanding of the dynamics of active particles in complex fluids, one should also consider such changes to the gait itself.

\section*{Acknowledgements}
Funding from the the Natural Sciences and Engineering Research Council of Canada (NSERC) is gratefully acknowledged. G.J.E. and C.D. also thank Professor G. M. Homsy for insightful discussions and support.
 
\bibliography{reference}

\end{document}